\documentclass[12pt,preprint]{aastex}

\usepackage{graphicx}             
\usepackage{pdflscape}
\usepackage{multirow}
\usepackage{array}
\usepackage{threeparttable}
\usepackage{booktabs}
\usepackage{color}
\usepackage{amssymb,amsmath}

\newcommand{\red}{}

\shorttitle{Seed population in SEPs and twin-CMEs}
\shortauthors{Ding, L.G., et al.}

\begin{document}


\title{Seed population in large Solar Energetic Particle events and the twin-CME scenario}


\author{{Liu-Guan Ding\altaffilmark{1}}, {Gang Li\altaffilmark{2,*}}, {Gui-Ming Le\altaffilmark{3}}, {Bin Gu\altaffilmark{1}},
{Xin-Xin Cao\altaffilmark{1}} }





\altaffiltext{1}{School of Physics and Optoelectronic Engineering,  Institue of Space Weather, Nanjing
University of Information Science \& Technology, Nanjing, Jiangsu, 210044, China}
\altaffiltext{2}{Department of Space Science and CSPAR, University of Alabama in
Huntsville, AL, 35899, USA *send correspondence to: gangli.uah@gmail.com}
\altaffiltext{3}{National Center for Space Weather, China Meteorological Administration, Beijing, 100081, China}


\begin{abstract}
It has been recently suggested that large solar energetic particle (SEP) events are often caused by twin CMEs.
In the twin-CME scenario, the preceding CME is to provide both an enhanced turbulence level and
enhanced seed population at the main CME-driven shock.
In this work, we study the effect of the preceding CMEs on the seed population.
We examine event-integrated abundance of iron to oxygen ratio (Fe/O) at energies above 25 MeV/nuc for large SEP events
in solar cycle 23.
We find that the Fe/O ratio (normalized to the reference coronal value of $0.134$) $\leq2.0$ for almost all single-CME
events and these events tend to have smaller peak intensities.
In comparison, the Fe/O ratio of twin-CME events scatters in a larger range, reaching as high as $8$,
suggesting the presence of flare material from perhaps preceding flares.
For extremely large SEP events with peak intensity above $1000$ pfu, the Fe/O drop below $2$, indicating that
in these extreme events the seed particles are dominated by coronal material than flare material.
The Fe/O ratios of Ground level enhancement (GLE) events, all being twin-CME events, scatter in a broad range.
For a given Fe/O ratio, GLE events tend to have larger peak intensities than non-GLE events. Using
velocity dispersion analysis (VDA), we find that GLE events have lower solar particle release (SPR)
heights than non-GLE events, \red{agreeing with earlier results by \citet{Reames09b}}.
\end{abstract}



\section{Introduction}
\label{sec.intro}
Solar energetic particles (SEPs) is a major concern of space physics and space weather.
The energy of SEPs in large SEP events, and in particular ground level enhancement (GLE) events,
can reach up to $\sim$GeV/nuc. These particles are believed to be produced at and near the Sun mainly
via two processes: solar flares and coronal mass ejections (CMEs).  Historically, SEP events
are classified as ``impulsive'' and ``gradual'' events depending on the duration of the associated
soft X-ray observations.
It was later used to refer to events where the particle acceleration process occurs at flares and
 CME-driven shocks, respectively \citep{Cane.etal86,Reames95,Reames99}.

The most intense SEP events are almost always associated with
fast and wide CMEs \citep{Kahler.etal84, Reames95,Kahler96,Gopalswamy.etal02,Cliver.etal04,
Tylka.etal05,Kahler.Vourlidas13}.
However, not all fast and wide CMEs lead to large SEP events \citep{Kahler96,Ding.etal13}.
An earlier study by \citet{Kahler96} showed that both the maximum energy and the intensity of
energetic particles in SEP events tend to correlate with shock speed. Later, \citet{Kahler.etal00} suggested
that the ambient energetic particle intensity prior to the event may be an important factor in causing a
large SEP event. These seed population may be from previous flare remanet materials
or ambient corona materials \citep{Mason.etal99,Mason.etal00,Gopalswamy.etal04,Li.etal12,Ding.etal13}.

\citet{Gopalswamy.etal04} noted first that there is a strong correlation between high
particle intensity events and the existence of preceding CMEs within $24$ hrs ahead of the primary CMEs.
 \citet{Gopalswamy.etal04} also noted that the SEP intensity showed poor correlation with the flare class.
\citet{Li.Zank05a} proposed that two consecutive CMEs may provide a favorable environment for particle acceleration.
This was refined in \citet{Li.etal12} as the twin-CME scenario. In \citet{Li.etal12}, the authors studied all
16 GLE events in solar cycle 23 and found that  there were always preceding CMEs within a short period ($9$ hrs) of
the main fast CMEs.  \citet{Li.etal12} proposed that when two CMEs erupt in sequence from the same or nearby
active regions (ARs) within a short period of time, the preceding CME-driven shock can enhance the
turbulence level at the main CME-driven shocks through Alfv\'en wave excitation, and increase the seed population
at the second shock through pre-acceleration at the preceding shock and/or flare. Both enhanced turbulence level
and enhanced seed population will favor a more efficient particle acceleration process at the second shock, leading
to large SEP or GLE events.

Extending the work of \citet{Li.etal12}, \citet{Ding.etal13} examined the twin-CME scenario against
all large SEP events and fast CME ($>900$km/s) having western hemisphere source regions in solar cycle 23.
They found that  61\% twin CMEs lead to large SEP events as compared to
only 29\% single fast CMEs leading to large SEP events. Furthermore, of all western large SEP events, 73\%
are twin CMEs. These findings support the proposal that twin-CMEs are responsible for large SEP events. For the
twin-CME scenario to work, the preceding CME can not be too far from the main CME. For if so, both the enhanced
turbulence level and the enhanced seed population may decay.
In the work of \citet{Ding.etal14}, the authors refined the time interval threshold on the identification of the
twin-CME scenario to be $13$ hrs.

The relative abundances of different elements in SEP events provide important clues about
the seed populations. For example, at $1$ to several MeV/nuc, impulsive events
often has a large $^3$He/$^4$He ratio than the solar wind since $^3$He is considered a tracer element
of solar flares \citep{Reames.etal90}. Flares also tend to have higher Fe/O ratio than corona.
A widely used coronal Fe/O ratio is $0.134$, a value derived by \citet{Reames98}, who added together all
particle counts in $49$ large solar particle events in the range $5$-$12$MeV/nuc to obtain this reference value.
For events of which the Fe/O ratio is considerably higher than the corona value,
 and/or the abundance ratio of $^3$He/$^4$He is higher than the solar wind value, one may expect that the seed
particles to be dominated by plasma that has been heated by the accompanying solar flares or preheated by
preceding flares. On the other hand, if the Fe/O ratio is close to the coronal value,
one may expect that the seed particle to be of normal coronal and/or solar wind material.

People have used different thresholds of Fe/O ratio to infer the presence or absence of impulsive flare material.
\citet{Reames.etal90} have used $1.7$ (normalized to the coronal value of $0.134$)
as the threshold for the enegy range of $1.9$-$2.8$ MeV/nuc.
Later, \citet{Reames.Ng04} suggested that the normalized values of Fe/O ratio in large ``impulsive'' events
to be $3.3$. More recently, \citet{Cane.etal06} used a normalized Fe/O ratio of $2.0$ as
the indication of flare-like material in the energy range of $25$-$80$ MeV/nuc.
In this work, we follow \citet{Cane.etal06} and use a Fe/O ratio of $2.0$ as our threshold for the presence of
flare material. \red{We use the same energy range of $25$-$80$ MeV/nuc as \citet{Cane.etal06}.
Note that, as shown in \citet{Tylka.etal05} (their figure 1), the Fe/O for two otherwise similar events can
differ substantially above about 10 MeV/nuc. \citet{Tylka.Lee06} suggested that this difference might be
 due to difference in shock obliquity. As argued by \citet{Tylka.Lee06}, the injection energy of
a quasi-perpendicular shock can be much larger than a quasi-parallel shock and consequently, the seed population
at a quasi-perpendicular shock can be more flare-like than a quasi-parallel shock. However, simulations by
\citet{Giacalone.Ellison00,Giacalone05,Giacalone05a} suggested that the injection energy may not depend
strongly on the shock obliquity.  In any events,
we point out that the Fe/O ratio above $10$ MeV/nuc can vary largely from event to event and this is the main source of
uncertainty for our analysis.}

Note that both flare material and corona material can be accelerated in the same gradual event
\citep[e.g.][]{Mason.etal99,Li.Zank05,Li.Mewaldt09,Li.etal12,Mewaldt.etal12}.
In many large gradual SEP events, a notable feature is the initial strong enhancement of Fe/O ratios,
sometimes reaching values of $1$, comparable to impulsive SEP events.
Such a strong enhancement at the beginning phase of a large SEP event has led many researchers
\citep[e.g.][]{Reames.etal90,Cliver96,Cane.etal03, Li.Zank05} to the examination of the so-called ``hybrid''
events, where particles accelerated at both flares and CME-driven shocks can contribute to the same SEP event.
Later, \citet{Mason.etal06} noted that when the Fe time-intensity profiles were compared to
those of O at a higher energy, the time profiles of Fe and O became very similar and yield Fe/O ratios that
do not vary substantially with time. \citet{Mason.etal06} concluded that the initial peak of Fe/O ratio
may be due to transport effect. In a follow up study,
\citet{Mason.etal12} examined the effect of transport on the temporal evolution of SEP intensities
for different heavy ions for $17$ large SEP events. By employing a numerical transport model
which includes the effects of pitch angle scattering, convection, adiabatic cooling, and magnetic focusing,
the authors modeled the time intensity profiles of H, He, O and Fe at several energies from $386$ keV/nuc
to $40$ MeV/nuc. The model calculation showed that the Fe/O ratio taking at different energies for Fe and O
show little time variation, confirming the earlier results of  \citet{Mason.etal06}. Furthermore, the scaling
of the energy is decided by the interplanetary turbulence. In another work,
using observations of $Wind$ and $Ulysses$, \citet{Tylka.etal13} also
demonstrated that the initial enhancements of Fe/O are better understood as a transport effect,
driven by the different mass-to-charge ratios of Fe and O.

While the instantaneous Fe/O ratio can vary largely as a function of time
in an SEP event due to the fact that the transport process is $Q/A$ dependent, the event-integrated Fe/O ratio
does not depend on the transport process and therefore
may still provide clues about the source of the seed population.
If both flare material and coronal material can serve as the seed population for gradual SEP events,
is there some dependence of the event intensity on the seed material? Do larger events tend to have more flare
material or coronal material?  In this paper, we address such questions using observations of
major large gradual SEP events in solar cycle 23. Since many large SEP events are twin-CME events, our study
are performed in the context of the twin-CME scenario.

Our paper is organized as follows: in section \ref{sec.data} we discuss the data selection and analysis procedure;
in section \ref{sec.result} we present our analysis results; and section \ref{sec.disc} contains the conclusion
and discussion.

\section{Data Selection}
\label{sec.data}

For large SEP events, we use the large proton event list from the NOAA
``Solar Proton Events Affecting the Earth Environment''
list (found at \url{http://www.swpc.noaa.gov/ftpdir/indices/SPE.txt})
in the years 1997 through 2006.
Note that in this list, a large proton event is defined with intensity $>10$ pfu (1 pfu=$1proton~cm^{-2} s^{-1} sr^{-1}$)
in the $>10$ MeV channel of the Geostationary Operational Environmental Satellites (GOES) instrument.
In order to identify the source active region of solar energetic particles clearly, we only consider the events
where the associated CMEs occur in the front-side hemisphere of the sun. SEP lists and GLE lists from
\citet{Ding.etal13,Ding.etal14} and \citet{Li.etal12} are also used as the major event lists in this study.
Since we are interested in the twin-CME scenario, the information about twin-CMEs of each SEP event in
\citet{Ding.etal13} and \citet{Ding.etal14} are also used. For the twin-CME events, we follow the same identification
criteria as \citet{Ding.etal14}.
%
%
The selected events and their properties are summarized in Table~\ref{table.1}.

The first column in Table~\ref{table.1} shows the event number. The second and the third column are the date and the
onset time of the SEP event. Column 4 and 5 are the onset time of the associated CME and its projection speed on the
sky plane (from catalog at  \url{http://cdaw.gsfc.nasa.gov/CME\_list/}). Column 6 is the NOAA active region (AR) number,
and ``?'' denotes event that has no NOAA AR number.
Column 7 is the location of the source active region at the onset time of CME eruption.
The associated flare classes and onset times are shown in column~8 and~9, respectively, where '-' denotes that no flares
are identified. Column 10 is the peak proton flux intensity in the $>10$MeV channel of GOES satellite of the SEP event.

Following \citet{Cane.etal06}, we use $25$-$80$MeV/nuc heavy ions (Fe and O)
detected by the ACE/SIS \citep{Stone.etal98} instrument in our study. As shown in \citet{Zank.etal00, Li.etal03},
the maximum particle energy at a CME-driven shock quickly drops as the shock propagates out. For a particle with
 energy of $25$-$80$MeV/nuc, it is likely accelerated near the Sun. So the Fe/O ratio in the
energy range of $25$-$80$MeV/nuc can largely reflect the near-sun seed population of the event.
The Fe/O ratio was calculated by integrating the time-averaged SEP spectra from $25$ to $80$ MeV/nuc
over the duration of the event.  The time intervals used for obtaining the Fe/O ratio for each SEP event are given
in column 11 and 12 in Table~\ref{table.1}.  Column 13 shows the event-integrated Fe/O ratio
(normalized to the value $0.134$) in the energy range $25$-$80$ MeV/nuc.
Column 14 and 15 are the solar energetic particle propagation path length ($L$) from near the sun to 1 AU and
the solar particle release height (SPR-H), respectively.
The path length and the SPR-H are inferred from the ACE/SIS data by using the
method of Velocity Dispersion Analysis (VDA) (e.g . \citep{Tylka.etal03,Reames09a}).  A symbol of '-' denotes that
there was not enough valid ACE/SIS data available or that the results are unreasonable.
For events where the SPR results are listed in \citet{Reames09a}, we obtained
similar results and have used the results as listed in \citet{Reames09a}.
Symbol `np' in the column 16 indicates that no preceding CMEs within 13~hrs ahead of the main CME was found.
These are  ``single-CME'' events. Symbol `*' in the column 17 denotes GLE events.

\section{Analyses and Results}
\label{sec.result}
For our analysis, we follow \citet{Ding.etal13} and categorize all SEP events to
two groups: ``single-CME'' SEP events and ``twin-CME'' SEP events.  \red{Note that the
preceding CMEs are often slower and narrower. If they are too slow, they may not drive a shock and consequently no particle
acceleration can take place and there is no enhanced seed population nor enhanced turbulence level at the shock driven by the
main CME. Therefore, in identifying the preceding CME, we require the
speed of the CME to be larger than $300$ km/sec so that the likelihood it drives a shock is larger. However, it has to be pointed out
that there is no guarantee of a shock at the preceding CME. Consequently, as noted by \citet{Ding.etal13},
some of the twin-CMEs we identified should be counted as single CMEs.}.

\red{In the case that the preceding CME does drive a shock and there is particle acceleration and so
enhanced seed population nor enhanced turbulence level, one may wonder which one, the seed population or the enhanced turbulence, is
the more important factor in leading to a large SEP event.  Such a question is hard to answer
as neither the seed population nor the turbulence strength near the Sun is presently available.
Perhaps in the very extreme events, both factors play a role.
Incidentally, we note that enhanced seed particles from preceding CMEs may also exist in impulsive SEP events
(see e. g. lists in \citep{Reames.etal14, Reames14, Reames.etal15}).  }

\subsection{Longitudinal dependance of Fe/O}
\label{subsec.lon}
\begin{figure}[htb]
    \centering
    \noindent \includegraphics[width=0.95\textwidth]{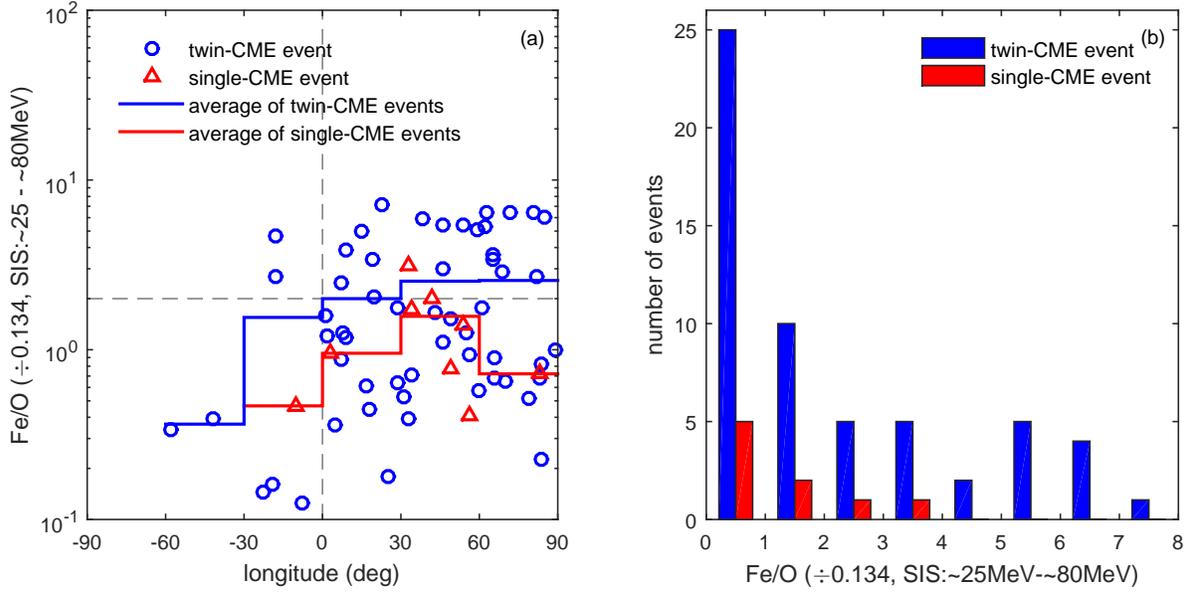}
    \caption{(a) Longitudinal distribution of the event-integrated Fe/O ratio in the energy range $25$-$80$ MeV/nuc (normalized to Reames value, 0.134) of SEP events. The blue circles denote the events produced by the twin-CME eruptions, and red triangles denote the events produced by the single-CME eruptions. The stair curves indicate the mean Fe/O values for each group in each longitude bin, where the blue ones represent the twin-CME events and the red ones represent  the single-CME events. (b) Histogram of the normalized Fe/O ratio of SEP events. The blue color show the twin-CME events, and the red color show the single-CME events. }
    \label{fig.feo_lon}
\end{figure}

For these two types of SEP events, we first examine the longitudinal distribution of the event-integrated Fe/O
(normalized to $0.134$) in the energy range of $25$-$80$MeV/nuc. This is shown in Figure~\ref{fig.feo_lon}.
In panel~(a) of Figure~\ref{fig.feo_lon}, the blue cycles denote ``twin-CME'' events,
and the red triangles denote ``single-CME'' events. From the figure we can see that, for both
``single-CME''  and ``twin-CME'' events, the Fe/O ratio for events with western source regions are
larger than those with eastern source region.
If we acknowledge that a normalized event-integrated Fe/O $>2.0$ to be an indication of
the presence of flare materials as the seed population
\citep{Cane.etal06}, then all, but two, events having flare material as seeds are from western hemisphere.
This is consistent with the case study of \citet{Reames14}, who noted that Fe-rich events tend to have
 shocks with western source.  Also shown in panel (a) are the blue and red stair curves which are
the mean Fe/O values in each bin of 30 degrees for the twin-CME events and the single-CME events,
respectively.  Clearly we can see that events with
high mean Fe/O values occur mainly in the longitude range from 30 to 90 degrees.
This is an interesting result. It suggests that the seed flare material may be very localized. We tend to
 see enhanced Fe/O ratio for western events because the magnetic connection to 1 AU is better for western events.
For eastern events, because the seed flare material are not well-connected to Earth, so we do not
see enhanced Fe/O ratio. This implies that if we observe large SEP events with multiple spacecraft that
are separated in longitude (e.g. STEREO spacecraft), we may observe different event-integrated Fe/O ratio.
Of course we do not know if the seed flare material is from the flare that accompanies the CME or from preceding
flares.

It is also interesting to note that, except two well-connected single-CME events, the rest Fe/O $>2.0$
events are all twin-CME events.
This is consistent with the twin-CME scenario in which the shock driven by the main CME not only can
accelerate coronal/solar wind material that are pre-accelerated at the shock with the preceding CME,
but also can accelerate flare material that are from flares associated with the preceding CMEs.
Preceding flares may provide seed particles for the subsequent acceleration at the main CME-driven shock
has been noted in the twin-CME scenario \citep{Li.etal12, Ding.etal13}.

We also note that for each longitude bin, the mean Fe/O ratio of twin-CME events is larger than that
of single-CME events. This may be related to an important question concerning large SEP events:
can particles accelerated at a flare be reaccelerated at the accompanying CME's driven shock?
If so, then the accompanying flare can provide seed
particles for the CME-driven shock. Such a scenario has been examined in the simulation of \citep{Li.Zank05}
where flare-accelerated particles are considered to be reaccelerated at the CME-driven shock.
However, our result that most Fe/O $>2.0$ events are twin-CME events seems to suggest that particles
 accelerated at a flare may have a hard time to be re-accelerated by the accompanying CME-driven shock.
Because if there is reacceleration, then single-CME event could have similar Fe/O ratio as twin-CME events.
On the other hand, if flares only provide seed particles for subsequent CME-driven shocks but not
the accompanying CME-driven shock, then we expect twin-CME events to have larger Fe/O ratio than
 single-CME events.

Panel (b) of Figure~\ref{fig.feo_lon} shows the histogram of the normalized event-integrated Fe/O ratio,
where the blue bars are for the twin-CME events, and the red ones are for the single-CME events.
While most single-CME events have the normalized Fe/O ratios below $2.0$,
a lot of twin-CME events have normalized Fe/O ratios between $2.0$ and $4.0$.
This again suggests that the seed particles in single-CME events are mostly from solar wind/coronal materials
while those in twin-CME events may have two sources: flare material from preceding flares and coronal/solar wind
material from preceding CMEs.

\subsection{SEP peak intensity and the normalized Fe/O ratio}
\label{subsec.inst}

\begin{figure}[htb]
    \centering
    \noindent \includegraphics[width=0.6\textwidth]{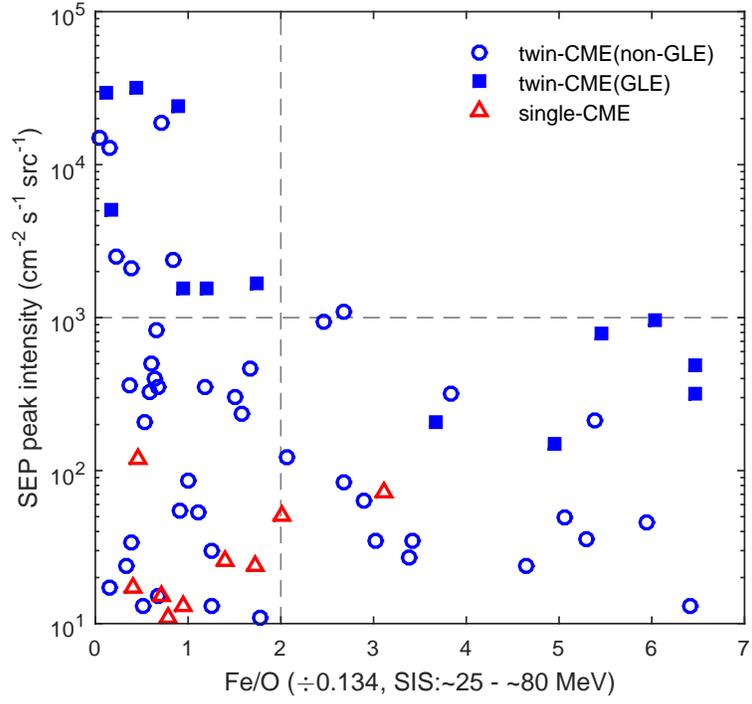}
    \caption{ Peak intensity of large SEP event plotted versus normalized Fe/O ratio. Blue and red Symbols indicate the large SEP events associated with twin-CME events and single-CME events respectively. The blue circles denote the non-GLE SEP events, and the blue squares denote the GLE events.
    }
    \label{fig.feo_inst}
\end{figure}

The peak flux intensities of large SEP events are shown in Figure~\ref{fig.feo_inst} as a function of the normalized
event-integrated Fe/O ratios. Blue symbols denote twin-CME events and red symbols denote single-CME events.
%
%
The result is very interesting. Consider first single-CME events.
These events are clustered in the lower left part of the plot. Except for one event,
all other $8$ events hav peak intensities smaller than 100 pfu. In the twin-CME scenario for large SEP events, the
peak intensity is directly related to the seed population. Lacking a preceding CME,  the seed population for
single-CME event is presumably limited and therefore we do not expect to find large SEP peak intensity for these events.
In comparison, twin-CME events show a much broader scatter of peak intensity: $14$ events have peak intensities
above $1000$ pfu;  $22$ have peak intensities between $100$ and $1000$ pfu, and $21$ have peak intensities between
$10$ and $100$ pfu. Such a scattering is a natural conclusion of the twin-CME scenario in which the preceding CMEs,
depending on its propagation direction, strength, how much time prior to the main CME, etc, can lead to a broad
range of enhanced seed population and enhanced turbulence level at the subsequent shock driven by the main CME.
\red{Unfortunately, no in-situ observations of either the seed population or the turbulence strength near the
Sun is presently available. We do not know which one,  the seed population or the enhanced turbulence, is
more important in leading to a large SEP event. Perhaps in the very extreme events, both factors play a role.}

One fact to note is that for extremely large SEP events, defined here as events with peak intensities
$>1000$pfu, all but $1$ (i.e. $13$ out of $14$) have normalized event-integrated Fe/O values $<2.0$, indicating
that the seed particles in these events are dominated by coronal material/solar wind.
As we discussed in the last section, flare material from preceding flares can serve as seed particles at the shock driven
by the main CME. So why extreme SEP events, which are all twin-CME events, and having preceding flares,
have Fe/O ratios smaller than $2.0$?  This can be understood if shocks driven by pre-CMEs can provide
more seed particles than pre-flares. This is possible because for flare seed particles (that are accelerated at the
pre-flares) to be later accelerated at the shock driven by the main CME, these flare particles need to
leak out to the interplanetary medium from the flare site first, through perhaps interchange reconnection. In comparisons,
coronal and solar wind material that are accelerated at the preceding shock can be processed by the main shock easily if
both shocks are propagating towards similar directions. Consequently, pre-CMEs may provide more seed particles than
pre-flares. Now an extremely large SEP event needs to have a large seed population, and since pre-CMEs can provide more
seed population than pre-flares, we conclude that extreme events are those events where seed particles are efficiently and
dominantly produced at pre-CMEs. Consequently the Fe/O ratios in extreme events are closer to coronal values than f
lare values. Along this reasoning, we would expect that in less extreme twin-CME events (e.g., events
having peak intensities smaller than $1000$ pfu), the Fe/O ratio will scatter and have a larger range. This is indeed
what is shown in Figure~\ref{fig.feo_inst}. Finally note that all single-CME events (except one) have small Fe/O ratios
since there is no contribution of flare seed material from pre-flares. A thumb rule of large SEP events
from Figure~\ref{fig.feo_inst} is therefore the following: a) single CMEs can rarely produce large SEP events with a
peak intensity larger than $100$ pfu and the Fe/O ratios in these events are often smaller than $2.0$;
b) large events having  peak intensity larger than $100$ pfu are almost all produced by twin-CMEs and the Fe/O ratio
of these events have a large scattering range, can reach $8.0$ if sufficient flare seed material presents;
3) for very large events having a peak intensity larger than $1000$ pfu, the seed population is likely
from pre-CMEs than pre-flares and therefore the Fe/O ratio becomes smaller again.

In the work of \citet{Ding.etal13}, it was suggested that if the twin-CME scenario is the cause for large SEP
events, then the peak flux intensity of these large SEP events (that are caused by twin-CMEs), which is mostly
controlled by the seed population and the turbulence level at the shock driven by the main CME, should have little
correlation with the associated flare class, or with the speed of the associated CMEs.
For extreme SEP events, the correlation should be even less.
Figure~\ref{fig.feo_flare_cme} plots the event peak intensity as functions of flare class and CME speed.
 The upper two panels are for all events and the lower two panels are for extreme events ($Ip>1000pfu$) only.
The blue dashed lines in each panel are the linear fit to the result and the correlation coefficients are shown
in each panel. Clearly, we can see from panel (c) and (d), for extreme SEP events there is no correlation between the
peak intensity and the flare class or the CME speed. Even for all events (panel (a) and (b)),  the dependence of peak
intensity on either the flare class or the CME speed is at most tangential. Figure~\ref{fig.feo_flare_cme},
 therefore, provides further supports to the twin-CME scenario.


%
%

Consider now GLE events.
In Figure~\ref{fig.feo_inst}, we also separate all twin-CME events to GLE events and non-GLE events.
The blue squares denote the GLE events, and the blue circles denote non-GLE events.
The red triangles denote single-CME events.
Not surprisingly, the GLE events are often the more intense events for a given Fe/O ratio.

\subsection{Solar Particle Release height}
\label{subsec.sprh}

We now examine the solar particle release (SPR) height for our events.
Using the VDA method \citep{Tylka.etal03,Reames09a} and data from ACE/SIS, we derive the Solar Particle Release (SPR) time near the
Sun. We then obtain the height of the associated CME from the speed of the CME
(as given from the CDAW CME catalog \url{http://cdaw.gsfc.nasa.gov/CME\_list/}).
 To be clear, it is the height of the nose of the CME and we use it as a proxy of the height of the
leading edge of the CME-driven shock. Note that only under the assumption that energetic particles are release near the nose of the
CME driven shock, this height is the solar particle release (SPR) height. If energetic particles are released at the flank of the
CME-driven shock, then this height is only the upper limit of the SPR height.
In the following we refer this height as the SPR height. This height is shown as a function of the Fe/O ratio
in Figure~\ref{fig.feo_sprh}.
As shown in the figure, it seems that events with higher Fe/O values have lower SPR heights. This can be understood as the following.
First we note that the seed particles from pre-CMEs can exist in a larger region (reaching to a larger height) than pre-flares since
the pre-CME is more extended that the pre-flare. Now since the SEPs are continuously accelerated at the shock driven by the main CME
and since the seed particles from pre-CMEs occupy a larger region than pre-flares, so the larger the SPR height,
the more pre-CME accelerated material (than pre-flare accelerated material) will be processed by the shock, thus a smaller Fe/O
 ratio. It is interesting to note that for GLE events, the SPR heights are low, $10$ of them are below $5 R_s$ and $3$
are between $5R_s$ and $10R_s$.
\red{This agrees with earlier work of \citet{Reames09b}, who first
suggested that GLE events have lower SPR release height than non-GLE events.}
This result indicates that to generate a GLE event, where particles can be accelerated to
$\sim$ GeV/nuc in energy, there must be plenty of seed particles, either flare material or corona material, due to preceding flares
and/or CMEs, to exist very low in the corona. This is not surprising since
coronal shock strength quickly decreases with height and the maximum particle energy also decreases quickly with CME
height \citep{Zank.etal00, Li.etal03}. So to get to $\sim$ GeV/nuc in energy, the acceleration has to occur very low in corona.

\section{Conclusions and Discussions}
\label{sec.disc}

In this work, we examine the source of seed population in large gradual SEP events using event-integrated Fe/O ratios
(normalized to the Reames value of $0.134$) in the energy range of $25$-$80$ MeV/nuc.
The data is from the SIS instrument  onboard ACE spacecraft are used.
We assume that the element abundances of Fe and O in this energy range is consistent with those in the seed population.
Following \citet{Cane.etal06} we use the normalized Fe/O ratio $> 2.0$ as the indicator of the presence of the flare materials.

We find that the SEP events with high Fe/O ($>2.0$) ratios almost always have western hemispheric source regions.
Furthermore all events with Fe/O $>2.0$ are twin-CME events except one, and  all single-CME events, except one,
have the normalized Fe/O ratio $\leq2.0$.
For twin-CME events, the normalized Fe/O ratio have a larger scattering, and many events have Fe/O ratios $>2.0$, indicating the
presence of flare seed material that are likely from pre-flares.
For any given longitude, the average Fe/O ratios in twin-CME events are higher than those in single-CME events.

These results are consistent with the two-CME scenario.
In the twin-CME scenario, the turbulence excited by preceding CME shocks can keep seed particles accelerated at either
preceding flares and preceding CME shocks near the Sun for a certain period, after which the seed particles can then undergo
a second acceleration process at the main CME shock. Consequently, because the possibility of the presence of flare seed particles,
 we expect to see enhancement of Fe/O in some of these twin-CME events. In comparison, we expect to see no F/O enhancement in
single-CME events.

\begin{figure}[htb]
    \centering
    \noindent \includegraphics[width=0.95\textwidth]{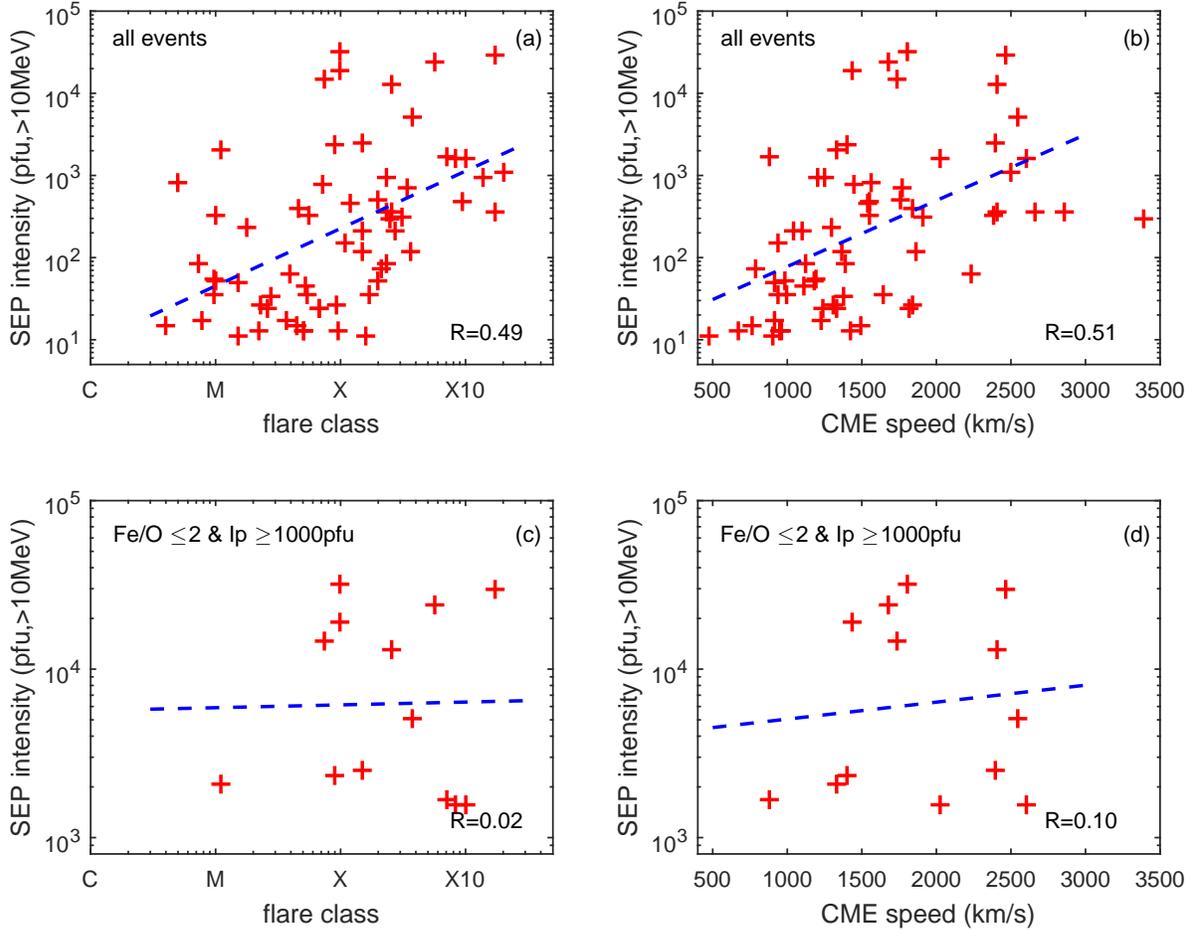}
    \caption{Peak intensity of large SEP event plotted versus the associated flare class and CME speed for all SEP events and extremely large SEP events. The blue line is the linear fit curve. }
    \label{fig.feo_flare_cme}
\end{figure}

\begin{figure}[htb]
    \centering
    \noindent \includegraphics[width=0.6\textwidth]{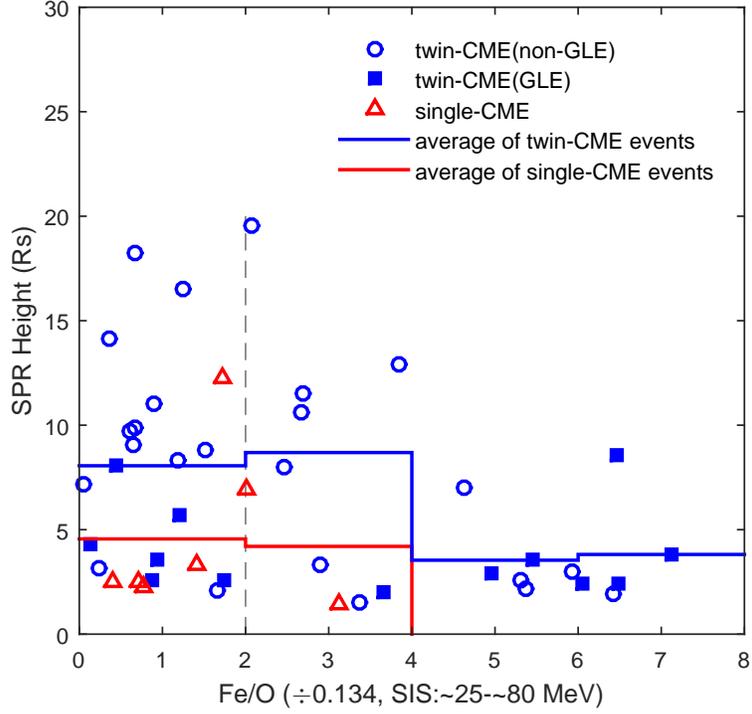}
    \caption{The solar particle release (SPR) height near the sun in each SEP event plotted versus the Fe/O ratio value. The blue and red symbols denote the twin-CME events and single-CME events respectively. The circles indicate the non-GLE SEPs, and the squares indicate the GLEs. The blue stairs show the mean SPR height in each Fe/O bin for twin-CME events, and the red stairs for single-CME events.
    }
    \label{fig.feo_sprh}
\end{figure}

It is interesting to find that for the most extremely large SEP events, defined as having $I_p>1000$ pfu, they are
all twin-CME events, but they all have Fe/O $\leq2.0$ except one.  We suggest that the reason for these events
to have Fe/O $\leq2.0$ is because to generate these events the intensity of the seed particles has to be large and comparing
to pre-flares, pre-CMEs can produce more seed particles, so the Fe/O ratio in these event are coronal like than flare like.
Furthermore the event-integrated intensity for extreme events show no correlation with the flare class or the CME speed
(as shown in Figure~\ref{fig.feo_flare_cme}).
This is predicted by the twin-CME scenario since in the twin-CME scenario the key factor to lead to a large SEP event is the
seed population and/or the presence of strong turbulence, but not the flare class nor the CME speed.

Not surprisingly the intensity of GLE events is usually larger than those of SEP events, suggesting the seed population for
GLE events is larger than non-GLE SEP events.
We also examine the solar particle release height for twin-CME events (GLE and non-GLE events) and single-CME events.
\red{We find that events with higher Fe/O values tend to have lower SPR heights, confirming earlier results of \citet{Reames09b}
for GLE events. }
This also agrees with the twin-CME scenario (see discussion in the last section).
Furthermore, the SPR height for GLE events are often lower than $5 R_s$, suggesting that to have a GLE event,
 plenty of seed particles must exist very low in the corona.

\acknowledgments
We are grateful to ACE/SIS, and CDAW CME catalogs for
making their data available online. This work is supported at UAH by NSF grant
AGS-1135432; 
at NUIST by NSFC-41304150 for Ding L.G.; at CMA by NSFC-41074132, NSFC-41274193 for Le G.M.;
and at NUIST by Jiangsu Government Scholarship for Overseas Studies (Grant No. JS2012-105) for Gu B..

\bibliographystyle{/home/ganli/PAPERS/BibStyles/apj.bst}



\clearpage

\begin{landscape}
\begin{table}[!htbp]
\caption{The properties of the large SEP events (solar cycle 23)}
\label{table.1}
\linespread{1.2}
\setlength{\tabcolsep}{4pt}
\scriptsize
\centering
\begin{tabular}{cccccccccccccccc}
\toprule
\multirow{2}{*}{No.}&  \multicolumn{2}{c}{SEP}   &CME  &$V_{CME}$& AR & \multirow{2}{*}{loc.}  & \multicolumn{2}{c}{ flare}
  &  Ip     & \multicolumn{2}{c}{summing interval$^a$}   & Fe/O$^b$ &$L^c$   &SPR-H$^d$ & \multirow{2}{*}{comm.$^e$}\\
\cline{2-3}
\cline{8-9}
\cline{11-12}
 &date&time&onset&(km/s)&(NOAA)&&FC&onset&(pfu)&start&end&(SIS)&(AU)&$(R_s)$ & \\
\hline
(1)&(2)&(3)&(4)&(5)&(6)&(7)&(8)&(9)&(10)&(11)&(12)&(13)&(14)&(15)&(16)\\
\midrule
1	&  1997/11/04	&  07:00	&  06:10	&  785	&   8100	&  S14W33	&  X2.1	&  05:52	&    72	&  11/04 06:00	&  11/06 09:00	&  3.12	  &  2.85	&    1.44	&	np 	\\
2	&  1997/11/06	&  13:00	&  12:11	&  1556	&   8100	&  S18W63	&  X9.4	&  11:49	&   490	&  11/06 12:00	&  11/09 20:00	&  6.47	  &  2.55	&    8.56	&	* 	\\
3	&  1998/05/02	&  14:00	&  14:06	&  938	&   8210	&  S15W15	&  X1.1	&  13:31	&   150	&  05/02 14:00	&  05/04 18:00	&  4.96	  &  1.24	&    2.90	&	*	\\
4	&  1998/05/06	&  08:25	&  08:29	&  1099	&   8210	&  S11W65	&  X2.7	&  07:58	&   210	&  05/06 08:00	&  05/08 05:00	&  3.66	  &  1.10	&    1.76	&	* 	\\
5	&  1999/06/04	&  08:00	&  07:27	&  2230	&   8552	&  N17W69	&  M3.9	&  06:52	&    64	&  06/04 07:00	&  06/06 04:00	&  2.89	  &  2.04	&    3.36	&	 	\\
6	&  2000/04/04	&  17:00	&  16:33	&  1188	&   8933	&  N16W66	&  C9.7	&  15:12	&    55	&  04/04 16:00	&  04/07 00:00	&  0.90	  &  1.66	&   11.04	&	 	\\
7	&  2000/06/10	&  18:00	&  17:08	&  1108	&   9026	&  N22W38	&  M5.2	&  16:40	&    46	&  06/10 17:00	&  06/12 03:00	&  5.94	  &  1.23	&    2.97	&	 	\\
8	&  2000/07/14	&  11:00	&  10:54	&  1674	&   9077	&  N22W07	&  X5.7	&  10:03	& 24000	&  07/14 11:00	&  07/18 19:00	&  0.89	  &  1.71	&    2.60	&	*	\\
9	&  2000/07/22	&  12:00	&  11:54	&  1230	&   9085	&  N14W56	&  M3.7	&  11:17	&    17	&  07/22 12:00	&  07/23 08:00	&  0.41	  &  1.78	&    2.48	&	np	\\
10	&  2000/09/12	&  13:00	&  11:54	&  1550	&   9163	&  S17W09	&  M1.0	&  11:30	&   320	&  09/12 13:00	&  09/15 23:00	&  3.84	  &  1.61	&   12.93	&	 	\\
11	&  2000/10/25	&  12:00	&  08:26	&  770	&   9199	&  N10W66	&  C4.0	&  08:45	&    15	&  10/25 13:00	&  10/27 21:00	&  0.68	  &  1.57	&    9.88	&	 	\\
12	&  2000/11/08	&  23:30	&  23:06	&  1738	&   9213	&  N10W77	&  M7.4	&  22:42	& 14800	&  11/08 23:00	&  11/13 00:00	&  0.05	  &  1.20	&    7.18	&	 	\\
13	&  2000/11/24	&  16:00	&  15:30	&  1245	&   9236	&  N22W07	&  X2.3	&  14:51	&   940	&  11/24 16:00	&  11/25 22:00	&  2.46	  &  1.07	&    8.03	&	 	\\
14	&  2001/01/28	&  17:00	&  15:54	&  916	&   9313	&  S04W59	&  M1.5	&  15:40	&    49	&  01/28 17:00	&  01/31 00:00	&  5.07	  &  	--	&  	--		&	 	\\
15	&  2001/03/29	&  11:00	&  10:26	&  942	&   9393	&  N20W19	&  X1.7	&  09:57	&    35	&  03/29 11:00	&  04/01 00:00	&  3.42	  &  	--	&  	--		&	 	\\
16	&  2001/04/02	&  23:00	&  22:06	&  2505	&   9393	&  N18W82	&  X20	&  21:32	&  1100	&  04/02 23:00	&  04/07 12:00	&  2.68	  &  1.11	&   10.63	&	 	\\
17	&  2001/04/10	&  08:00	&  05:30	&  2411	&   9415	&  S23W09	&  X2.3	&  05:06	&   355	&  04/10 08:00	&  04/12 12:00	&  1.18	  &  2.52	&    8.28	&	 	\\
18	&  2001/04/12	&  12:00	&  10:31	&  1184	&   9415	&  S19W42	&  X2.0	&  09:39	&    51	&  04/12 12:00	&  04/14 00:00	&  2.01	  &  2.11	&    6.96	&	np 	\\
19	&  2001/04/15	&  14:00	&  14:07	&  1199	&   9415	&  S20W85	&  X14	&  13:19	&   951	&  04/15 14:00	&  04/18 00:00	&  6.04	  &  1.70	&    2.4	&	*	\\
20	&  2001/09/15	&  12:00	&  11:54	&  478	&   9608	&  S21W49	&  M1.5	&  11:04	&    11	&  09/15 12:00	&  09/16 06:00	&  0.78	  &  2.30	&    2.28	&	np 	\\
21	&  2001/10/01	&  13:00	&  05:30	&  1405	&   9628	&  S20W84	&  M9.1	&  04:41	&  2360	&  10/01 13:00	&  10/03 06:00	&  0.83	  &  --	&  	--		&	 	\\
22	&  2001/10/19	&  17:30	&  16:50	&  901	&   9661	&  N15W29	&  X1.6	&  16:13	&    11	&  10/19 18:00	&  10/21 20:00	&  1.77	  &  --	&    --		&	 	\\
23	&  2001/11/04	&  17:00	&  16:35	&  1810	&   9684	&  N06W18	&  X1.0	&  16:03	& 31700	&  11/04 17:00	&  11/09 13:00	&  0.44	  &  1.95	&    8.06	&	* 	\\
24	&  2001/11/23	&  00:00	&  23:30	&  1437	&   9704	&  S15W34	&  M9.9	&  22:32	& 18900	&  11/23 00:00	&  11/26 23:00	&  0.71	  &  --	&  	--		&	 	\\
25	&  2001/12/26	&  05:30	&  05:30	&  1446	&   9742	&  N08W54	&  M7.1	&  04:32	&   779	&  12/26 06:00	&  12/28 22:00	&  5.46	  &  1.64	&    3.6	&	* 	\\
26	&  2002/01/14	&  06:00	&  05:35	&  1492	&      ?	&  S28W83	&  M4.4	&  05:29	&    15	&  01/14 06:00	&  01/17 00:00	&  0.72	  &  1.70	&    2.46	&	np 	\\
27	&  2002/02/20	&  06:30	&  06:30	&  952	&   9825	&  N12W72	&  M5.1	&  05:52	&    13	&  02/20 06:00	&  02/21 12:00	&  6.42	  &  1.47	&    1.91	&	 	\\
28	&  2002/03/16	&  03:00	&  23:06	&  957	&   9866	&  S08W03	&  M2.2	&  22:09	&    13	&  03/16 03:00	&  03/17 16:00	&  0.95	  &  	--	& 	--	 	&	np 	\\
29	&  2002/03/18	&  06:00	&  02:54	&  989	&   9866	&  S09W46	&  M1.0	&  02:16	&    53	&  03/18 06:00	&  03/20 09:00	&  1.10	  &  	--	&  	--		&	 	\\
30	&  2002/04/17	&  10:30	&  08:26	&  1240	&   9906	&  S14W34	&  M2.6	&  07:46	&    24	&  04/17 11:00	&  04/18 20:00	&  1.72	  &  1.31	&   12.23	&	np 	\\
31	&  2002/04/21	&  02:00	&  01:27	&  2393	&   9906	&  S14W84	&  X1.5	&  00:43	&  2520	&  04/21 02:00	&  04/25 00:00	&  0.23	  &  1.68	&    3.12	&	 	\\
32	&  2002/05/22	&  06:00	&  03:50	&  1557	&   9948	&  S15W70	&  C5.0	&  03:18	&   820	&  05/22 06:00	&  05/25 00:00	&  0.65	  & 	--	&  	--		&	 	\\
33	&  2002/07/16	&  10:30	&  21:30	&  1300	&  10030	&  N19W01	&  M1.8	&  21:03	&   234	&  07/16 11:00	&  07/19 08:00	&  1.59	  &  	--	&  	--		&	 	\\
34	&  2002/08/14	&  03:00	&  02:30	&  1309	&  10061	&  N09W54	&  M2.3	&  01:47	&    26	&  08/14 01:00	&  08/15 22:00	&  1.40	  &  1.35	&    3.33	&	np 	\\
35	&  2002/08/22	&  02:30	&  02:06	&  998	&  10069	&  S07W62	&  M5.4	&  01:47	&    36	&  08/22 02:00	&  08/23 06:00	&  5.30	  &  1.69&    2.62	&		 \\
36	&  2002/08/24	&  01:30	&  01:27	&  1913	&  10069	&  S02W81	&  X3.1	&  00:49	&   317	&  08/24 02:00	&  08/27 00:00	&  6.47	  &  2.16&    2.4	&	*	 \\
\bottomrule
\end{tabular}
\end{table}
\end{landscape}

\begin{landscape}
\begin{table}[!htbp]
{\centering{{\bf Table~\ref{table.1}} (Continued)}}\\
\linespread{1.2}
\setlength{\tabcolsep}{4pt}
\scriptsize
\centering
\begin{tabular}{cccccccccccccccc}
\toprule
\multirow{2}{*}{No.}&  \multicolumn{2}{c}{SEP}   &CME  &$V_{CME}$& AR & \multirow{2}{*}{loc.}  & \multicolumn{2}{c}{ flare}
  &  Ip     & \multicolumn{2}{c}{summing interval$^a$}   & Fe/O$^b$  &$L^c$   &SPR-H$^d$ & \multirow{2}{*}{comm.$^e$}\\
\cline{2-3}
\cline{8-9}
\cline{11-12}
 &date&time&onset&(km/s)&(NOAA)&&FC&onset&(pfu)&start&end&(SIS)&(AU)&$(R_s)$ & \\
\hline
(1)&(2)&(3)&(4)&(5)&(6)&(7)&(8)&(9)&(10)&(11)&(12)&(13)&(14)&(15)&(16)\\
\midrule

37	&  2002/09/06	&  04:00	&  02:06	&  737	&  10095	&  N08W31	&  --	&      --	&   208	&  09/06 04:00	&  09/09 00:00	&  0.53   &  	--	&  	--		&		 \\
38	&  2002/11/09	&  15:00	&  13:32	&  1838	&  10180	&  S12W29	&  M4.6	&  13:08	&   404	&  11/09 15:00	&  11/11 12:00	&  0.64   &  2.20&    9.09	&		 \\
39	&  2003/05/28	&  04:00	&  00:50	&  1366	&  10365	&  S07W20	&  X3.6	&  00:17	&   121	&  05/28 04:00	&  05/30 03:00	&  2.07   &  2.24&   19.54	&		 \\
40	&  2003/05/31	&  02:40	&  02:30	&  1835	&  10365	&  S07W65	&  M9.3	&  02:13	&    27	&  05/31 03:00	&  06/01 00:00	&  3.37   &  1.85&    1.54	&		 \\
41	&  2003/10/26	&  18:25	&  17:54	&  1537	&  10484	&  N04W43	&  X1.2	&  17:21	&   466	&  10/26 17:00	&  10/27 22:00	&  1.67   &  1.28&    2.10	&		 \\
42	&  2003/10/29	&  21:00	&  20:54	&  2029	&  10486	&  S15W02	&  X10	&  20:37	&  1570	&  10/29 21:00	&  11/01 12:00	&  1.21   &  1.75&    5.7	&	*	 \\
43	&  2003/11/02	&  11:00	&  09:30	&  2036	&  10486	&  S17W55	&  --	&      --	&    30	&  11/02 10:00	&  11/02 17:00	&  1.25   &  1.42&   16.55	&		 \\
44	&  2003/11/02	&  18:00	&  17:30	&  2598	&  10486	&  S14W56	&  X8.3	&  17:03	&  1570	&  11/02 18:00	&  11/04 20:00	&  0.94   &  1.83&    3.57	&	*	 \\
45	&  2003/11/04	&  22:25	&  19:54	&  2657	&  10486	&  S19W83	&  X17	&  19:29	&   353	&  11/04 23:00	&  11/07 06:00	&  0.68   &  2.41&   18.27	&		 \\
46	&  2003/11/20	&  08:30	&  08:06	&  669	&  10501	&  N01W08	&  M9.6	&  07:35	&    13	&  11/20 08:00	&  11/21 12:00	&  1.26   & 	-- 	&  	--		&		 \\
47	&  2003/12/02	&  12:30	&  10:50	&  1393	&  10508	&  S19W89	&  C7.2	&  09:40	&    86	&  12/02 13:00	&  12/04 10:00	&  1.00   & 	--	& 	--	 	&		 \\
48	&  2004/04/11	&  06:00	&  04:30	&  1645	&  10588	&  S16W46	&  C9.6	&  03:54	&    35	&  04/11 06:00	&  04/12 20:00	&  3.02   &  	--	&  	--		&		 \\
49	&  2004/07/25	&  18:55	&  14:54	&  1333	&  10652	&  N08W33	&  M1.1	&  14:19	&  2086	&  07/25 18:00	&  07/28 13:00	&  0.39   & 	--	&  	--		&		 \\
50	&  2004/11/07	&  18:00	&  16:54	&  1759	&  10696	&  N09W17	&  X2.0	&  15:42	&   495	&  11/07 18:00	&  11/09 16:00	&  0.61   &  1.89&    9.75	&		 \\
51	&  2004/11/10	&  03:00	&  02:26	&  3387	&  10696	&  N09W49	&  X2.5	&  01:59	&   300	&  11/10 03:00	&  11/15 00:00	&  1.51   &  1.61&    8.85	&		 \\
52	&  2005/01/16	&  00:10	&  23:07	&  2861	&  10720	&  N15W05	&  X2.6	&  22:25	&   365	&  01/16 00:00	&  01/17 09:00	&  0.36   &  1.91&   14.10	&		 \\
53	&  2005/01/17	&  13:05	&  09:54	&  2547	&  10720	&  N15W25	&  X3.8	&  09:42	&  5040	&  01/17 13:00	&  01/19 08:00	&  0.18   &  -- &   --	&	*	 \\
54	&  2005/01/20	&  07:00	&  06:54	&  882	&  10720	&  N14W61	&  X7.1	&  06:36	&  1680	&  01/20 07:00	&  01/22 00:00	&  1.75   &  1.19&    2.6	&	*	 \\
55	&  2005/07/13	&  17:00	&  14:30	&  1423	&  10786	&  N08W79	&  M5.0	&  14:01	&    13	&  07/13 17:00	&  07/14 10:00	&  0.51   &  	--	&  	--		&		 \\
56	&  2005/08/22	&  19:30	&  17:30	&  2378	&  10798	&  S12W60	&  M5.6	&  16:46	&   330	&  08/22 20:00	&  08/26 00:00	&  0.58   &  	--	&  	--		&		 \\
57	&  2006/12/13	&  03:10	&  02:54	&  1774	&  10930	&  S06W23	&  X3.4	&  02:14	&   698	&  12/13 03:00	&  12/14 20:00	&  7.12   &  1.72&    3.8	&	*	 \\
58	&  2006/12/14	&  22:55	&  22:30	&  1042	&  10930	&  S06W46	&  X1.5	&  21:07	&   215	&  12/14 23:00	&  12/16 18:00	&  5.38   &  1.77&    2.14	&		 \\
E1	&  2000/06/06	&  19:00	&  15:54	&  1119	&   9026	&  N20E18	&  X2.3	&  14:58	&    84	&  06/06 22:00	&  06/10 00:00	&  2.68   &  2.15&   11.49	&		 \\
E2	&  2001/08/09	&  19:00	&  21:30	&  909	&   9570	&  S17E19	&  C7.8	&  18:22	&    17	&  08/09 19:00	&  08/11 12:00	&  0.16   & 	-- 	&  	--		&		 \\
E3	&  2001/09/24	&  11:00	&  10:31	&  2402	&   9632	&  S16E23	&  X2.6	&  09:32	& 12900	&  09/24 11:00	&  10/01 00:00	&  0.14   &  	--	& 	--	 	&		 \\
E4	&  2001/10/22	&  17:00	&  15:06	&  1336	&   9672	&  S21E18	&  M6.7	&  14:27	&    24	&  10/22 17:00	&  10/26 01:00	&  4.64   &  2.58&    7.00	&		 \\
E5	&  2001/11/17	&  06:00	&  05:30	&  1379	&   9704	&  S13E42	&  M2.8	&  04:49	&    34	&  11/17 12:00	&  11/22 12:00	&  0.39   & 	--	&  	--		&		 \\
E6	&  2003/06/18	&  10:00	&  23:18	&  1813	&  10386	&  S08E58	&  M6.8	&  22:27	&    24	&  06/18 10:00	&  06/20 18:00	&  0.34   &  --	&  --		&		 \\
E7	&  2003/10/28	&  12:00	&  11:30	&  2459	&  10486	&  S16E08	&  X17	&  11:00	& 29500	&  10/28 11:00	&  10/29 20:00	&  0.13   &  1.38&    4.3	&	*	 \\
E8	&  2005/09/13	&  23:30	&  20:00	&  1866	&  10808	&  S09E10	&  X1.5	&  19:19	&   120	&  09/13 23:00	&  09/16 22:00	&  0.47   & 	--	& 	--	 	&	np	 \\
\bottomrule
\end{tabular}
\begin{itemize}
\scriptsize
  \item[$a$] the time interval of calculating the event-integrated Fe/O;
  \item[$b$] the normalized value of event-integrated Fe/O in the range of $25$-$80$MeV (SIS) (normalized to Reames value 0.134);
  \item[$c$] the path length of SEP propagating from the sun to the spacecraft deduced by the VDA method; `-' denote that there is no credible and reasonable results, or no enough valid ACE/SIS data;
  \item[$d$] the solar particle release height of SEP deduced by VDA and CME speed, and `-' denote that there is no credible and reasonable results, or no enough valid ACE/SIS data;
  \item[$e$] `*' denote the GLE events, and `np' denote the SEP events (called single-CME events) that have no identified preceding CMEs within 13 hrs ahead of the associated fast and wide CME.
\end{itemize}
\end{table}
\end{landscape}

\label{lastpage}

\end{document}